\documentclass[runningheads]{llncs}
\usepackage{amsmath}
\usepackage{amssymb}

\usepackage[T1]{fontenc}
\usepackage{graphicx}
\usepackage{tikz}
\usetikzlibrary{3d, perspective}
\usepackage{breakurl}
\usepackage{tcolorbox}
\usepackage{float}
\usepackage{subcaption}

\newcommand\frcap{\mathrm{fcap}}

\newcommand{\ens}[1][e] {\mathsf{#1}} % Ensemble
\newcommand{\Ens}[1][E] {\mathcal{#1}} % Ensemble space

\def\>{\rangle}
\def\<{\langle}

\begin{document}
\title{What non-additive integral for ensemble spaces?}
\author{Gabriele Carcassi\inst{1}\orcidID{0000-0002-1071-6251}, \\
Christine A. Aidala\inst{1}\orcidID{0000-0001-9540-4988} \\ 
\and Tobias Thrien\inst{1}\orcidID{0009-0005-9232-2617}}
\authorrunning{G. Carcassi, C. A. Aidala, et al.}
\institute{University of Michigan, Ann Arbor, MI 48109, US \\
\email{\{carcassi, caidala, thrien\}@umich.edu}}
\maketitle              % typeset the header of the contribution
\begin{abstract}
In a previous work we were able to define a non-additive measure that can be used to represent both classical and quantum states in physics. We further extended this idea to work on a generic space of statistical ensembles (i.e. an ensemble space) in a way that connects to Choquet theory. The question of which non-additive integral is suitable to generalize the notion of expectation value remains open. In this paper we show that the Sugeno and Choquet integrals are not suitable.

\keywords{Ensemble spaces  \and Choquet theory \and Choquet integral.}
\end{abstract}
\section{Introduction}

In a previous work \cite{aop-phys-NonAdditiveMeasuresForQuantum} we introduced our effort to use non-additive measures to represent states in physical theories as a way to address the non-Kolmogorovian nature of probability in quantum mechanics \cite{gleason1957measures,groenewold1946principles,gudder2009quantum,hamhalter2003quantum,moyal1949quantum,sorkin1994quantum,svozil2022extending}. This has led us to work on a general theory for statistical ensembles, which can serve as a common base theory for any physical theory, including classical mechanics and quantum mechanics. An ensemble space can be shown to be a bounded convex subset of a topological vector space, and each element can be represented by a non-additive measure that we call fraction capacity. To give some intuition, in the finite-dimensional case, this connects very nicely to Choquet theory as the fraction capacity is simply the supremum of all the measures that can represent the same element.

It is still an open question what correct notion of non-additive integral recovers the expectation of a random variable. In this work we show that the two main standard integrals, the Sugeno and Choquet integrals, do not work for our purposes. The ultimate goal is to engage with experts in the non-additive/fuzzy measure community to find what integral does work.

\section{Choquet theory and fraction capacity}

\begin{figure}[h]
	\centering
	
	% perspective
	\newcommand{\yaw}{11}
	\newcommand{\pitch}{28}
	\newcommand{\roll}{23}
	
	% transformation matrix = components of data axes in canvas coordinates
	\pgfmathsetmacro{\xx}{cos(\yaw)*cos(\pitch)}
	\pgfmathsetmacro{\xy}{sin(\yaw)*cos(\pitch)}
	\pgfmathsetmacro{\xz}{-sin(\pitch)}
	
	\pgfmathsetmacro{\yx}{cos(\yaw)*sin(\pitch)*sin(\roll) - sin(\yaw)*cos(\roll)}
	\pgfmathsetmacro{\yy}{sin(\yaw)*sin(\pitch)*sin(\roll) + cos(\yaw)*cos(\roll)}
	\pgfmathsetmacro{\yz}{cos(\pitch)*sin(\roll)}
	
	\pgfmathsetmacro{\zx}{cos(\yaw)*sin(\pitch)*cos(\roll) + sin(\yaw)*sin(\roll)}
	\pgfmathsetmacro{\zy}{sin(\yaw)*sin(\pitch)*cos(\roll) - cos(\yaw)*sin(\roll)}
	\pgfmathsetmacro{\zz}{cos(\pitch)*cos(\roll)}
	
	% intersection of yz grand circle with horizon
	\pgfmathsetmacro{\ixx}{0}
	\pgfmathsetmacro{\ixy}{\zz / sqrt(\zz*\zz + \yz*\yz)}
	\pgfmathsetmacro{\ixz}{-\yz / sqrt(\zz*\zz + \yz*\yz)}
	% intersection of zx grand circle with horizon
	\pgfmathsetmacro{\iyx}{\zz / sqrt(\zz*\zz + \xz*\xz)}
	\pgfmathsetmacro{\iyy}{0}
	\pgfmathsetmacro{\iyz}{-\xz / sqrt(\zz*\zz + \xz*\xz)}
	% intersection of xy grand circle with horizon
	\pgfmathsetmacro{\izx}{-\yz / sqrt(\xz*\xz + \yz*\yz)}
	\pgfmathsetmacro{\izy}{\xz / sqrt(\xz*\xz + \yz*\yz)}
	\pgfmathsetmacro{\izz}{0}
	
	% dimensions
	\newcommand{\radius}{1}
	
	\begin{tikzpicture}[x={(\xx cm, \xy cm)}, y={(\yx cm, \yy cm)}, z={(\zx cm, \zy cm)}, scale=1.2]
		% Triangle
		\draw (canvas cs: x=-3cm, y=-0.25cm) +(canvas polar cs: angle=90, radius=1cm) -- +(canvas polar cs: angle=210, radius=1cm) -- +(canvas polar cs: angle=330, radius=1cm) -- cycle;
		
		% Bloch sphere (outline drawn in canvas plane)
		\draw (0, 0, 0) circle (\radius cm);
		
		% meridians and equator
		\begin{scope}[canvas is yz plane at x=0]
			\draw (\ixy, \ixz) arc[start angle={atan2(\ixz, \ixy)}, delta angle=180, radius=1 cm];
			\draw[dashed] (-\ixy, -\ixz) arc[start angle={atan2(-\ixz, -\ixy)}, delta angle=180, radius=1 cm];
		\end{scope}
		\begin{scope}[canvas is zx plane at y=0]
			\draw[dashed] (\iyz, \iyx) arc[start angle={atan2(\iyx, \iyz)}, delta angle=180, radius=1 cm];
			\draw (-\iyz, -\iyx) arc[start angle={atan2(-\iyx, -\iyz)}, delta angle=180, radius=1 cm];
		\end{scope}
		\begin{scope}[canvas is xy plane at z=0]
			\draw[dashed] (\izx, \izy) arc[start angle={atan2(\izy, \izx)}, delta angle=180, radius=1 cm];
			\draw (-\izx, -\izy) arc[start angle={atan2(-\izy, -\izx)}, delta angle=180, radius=1 cm];
		\end{scope}
	\end{tikzpicture}
	\caption{Examples of a classical discrete ensemble space over three elements (triangle) and of a two-dimensional quantum ensemble space (ball).}\label{convex_examples}
\end{figure}
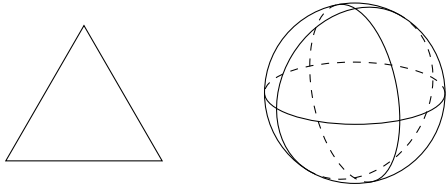

To simplify the problem, let us assume that an ensemble space $\Ens$ is a compact convex subset of a finite-dimensional vector space. Our two main ensembles, as exemplified in Fig. \ref{convex_examples}, are the space of classical probability measures with a finite sample space (i.e. an $n$-dimensional simplex) and the space of quantum pure and mixed states (i.e. the set of density matrices for an $n$-dimensional quantum system). This is somewhat restrictive, as classical mechanics is infinite-dimensional, but it will suffice for our purposes.

Given an element $\ens \in \Ens$, the \textbf{fraction capacity} of $A \subseteq \Ens$ with respect to $\ens$ is defined as
\begin{equation}
	\frcap_{\ens}(A) = \sup(\{ p \in [0,1] \, | \, \exists \, \ens[a] \in A, \ens[b] \in \Ens \text{ s.t. }  \ens = p \ens[a] + \bar{p} \ens[b] \}),
\end{equation}
which returns the biggest fraction of $\ens$ that can be expressed as a convex combination of elements from $A$. This is a monotone, subadditive set function that is continuous from below, and its restriction $\frcap_{\ens}|_X$ over the extreme points $X$ (i.e. the pure ensembles) uniquely defines a given $\ens$.

Note that, since we assumed that $\Ens$ is compact and convex, Choquet theory applies. In this case, the fraction capacity is simply the supremum of all possible probability measures that represent $\ens$. Notably, the fraction capacity is additive if and only if $\ens$ is represented only by a single probability measure, which means $\Ens$ is a simplex. This gives a simple and intuitive characterizations of classical probability spaces in terms of additivity of the capacities.

\section{Generalizing expectations}

A \textbf{statistical variable} is an affine function $F : \Ens \to \mathbb{R}$ as it returns the expectation for each ensemble. In the classical case, since $\frcap_{\ens}|_X$ is additive, we simply have
\begin{equation}
	E[F \, | \, \ens] = F(\ens) = \int_X F(x) d\frcap_{\ens}|_X.
\end{equation}
The goal, then, is to find a non-additive integral $(E)\int_X$ such that
\begin{equation}
	E[F \, | \, \ens] = F(\ens) = (E)\int_X F(x) d\frcap_{\ens}|_X
\end{equation}
which reduces to the Riemann integral in the classical (additive) case. The integral will depend only on the values of $F$ and $\frcap_{\ens}|_X$, but since these uniquely determine both $F$ and $\ens$, the integral can be understood as a map from $F$ and $\ens$ to $F(\ens)$, which is well-defined. Therefore, in principle, this integral should exist.

\section{Non-suitability of Sugeno and Choquet integrals}

As there are different notions of non-additive measures and integrals \cite{denneberg1994non,grabisch2016,pap2013,nonadditive2014}, our first step was to see whether the two most common non-additive integrals, the Sugeno integral and the Choquet integral, are suitable.

The Sugeno integral does not recover the Lebesgue integral in the case that the capacity is additive (i.e.~classical), and is therefore quickly ruled out. The Choquet integral does, so it will work in the classical case. However, it does not work in the quantum case. We can show this by computing the Choquet integral for a two-state quantum system and see that it does not recover the expectation of the statistical variables in the corresponding ensemble space.

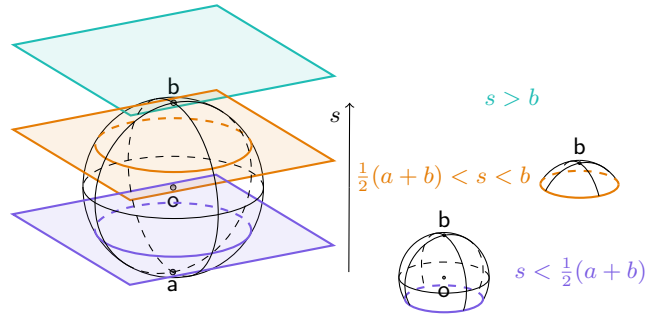
\begin{figure}[h]
	\centering
	
	% perspective
	\newcommand{\yaw}{11}
	\newcommand{\pitch}{28}
	\newcommand{\roll}{23}
	
	% transformation matrix = components of data axes in canvas coordinates
	\pgfmathsetmacro{\xx}{cos(\yaw)*cos(\pitch)}
	\pgfmathsetmacro{\xy}{sin(\yaw)*cos(\pitch)}
	\pgfmathsetmacro{\xz}{-sin(\pitch)}

	\pgfmathsetmacro{\yx}{cos(\yaw)*sin(\pitch)*sin(\roll) - sin(\yaw)*cos(\roll)}
	\pgfmathsetmacro{\yy}{sin(\yaw)*sin(\pitch)*sin(\roll) + cos(\yaw)*cos(\roll)}
	\pgfmathsetmacro{\yz}{cos(\pitch)*sin(\roll)}
	
	\pgfmathsetmacro{\zx}{cos(\yaw)*sin(\pitch)*cos(\roll) + sin(\yaw)*sin(\roll)}
	\pgfmathsetmacro{\zy}{sin(\yaw)*sin(\pitch)*cos(\roll) - cos(\yaw)*sin(\roll)}
	\pgfmathsetmacro{\zz}{cos(\pitch)*cos(\roll)}

	% intersection of yz grand circle with horizon
	\pgfmathsetmacro{\ixx}{0}
	\pgfmathsetmacro{\ixy}{\zz / sqrt(\zz*\zz + \yz*\yz)}
	\pgfmathsetmacro{\ixz}{-\yz / sqrt(\zz*\zz + \yz*\yz)}
	% intersection of zx grand circle with horizon
	\pgfmathsetmacro{\iyx}{\zz / sqrt(\zz*\zz + \xz*\xz)}
	\pgfmathsetmacro{\iyy}{0}
	\pgfmathsetmacro{\iyz}{-\xz / sqrt(\zz*\zz + \xz*\xz)}
	% intersection of xy grand circle with horizon
	\pgfmathsetmacro{\izx}{-\yz / sqrt(\xz*\xz + \yz*\yz)}
	\pgfmathsetmacro{\izy}{\xz / sqrt(\xz*\xz + \yz*\yz)}
	\pgfmathsetmacro{\izz}{0}

	% dimensions
	\newcommand{\s}{-0.5}
	\pgfmathsetmacro{\radiusats}{sqrt(1 - \s*\s)}
	\pgfmathsetmacro{\stheta}{asin(\s)}
	\pgfmathsetmacro{\minstheta}{asin(-\s)}
	
	\newcommand{\radius}{1}
	\newcommand{\margin}{1.3}
	
	\definecolor{colora}{HTML}{1EBCB4}
	\definecolor{colorb}{HTML}{E57C04}
	\definecolor{colorc}{HTML}{7A5CE3}
	
	\begin{tikzpicture}[x={(\xx cm, \xy cm)}, y={(\yx cm, \yy cm)}, z={(\zx cm, \zy cm)}, scale=1.2]
		% Bloch sphere (outline drawn in canvas plane)
		\draw (0, 0, 0) circle (\radius cm);
		
		% special points
		\shadedraw (0, 0, 0) circle (0.03 cm) node[below] {$\mathsf{o}$};
		\shadedraw (0, \radius, 0) circle (0.03 cm) node[above] {$\mathsf{b}$};
		\shadedraw (0, -\radius, 0) circle (0.03cm) node[below] {$\mathsf{a}$};
		
		% s axis
		\begin{scope}[shift={(1.95*\xx, 1.95*\yx, 1.95*\zx)}]
			\draw (0, -1, 0) -- (0, 0, 0);
			\draw[->] (0, 0, 0) -- (0, 1, 0) node[below left] {$s$};
		\end{scope}
		
		% cut 1
		\begin{scope}[canvas is zx plane at y=\radius-\s]
			\filldraw[color=colora, thick, draw opacity=1, fill opacity=0.1] (-\margin, -\margin) -- (\margin, -\margin) -- (\margin, \margin) -- (-\margin, \margin) -- cycle;
		\end{scope}

		% cut 2
		\begin{scope}[canvas is zx plane at y=-\s]
			\filldraw[color=colorb, thick, draw opacity=1, fill opacity=0.1] (-\margin, -\margin) -- (\margin, -\margin) -- (\margin, \margin) -- (-\margin, \margin) -- cycle;
			\draw[color=colorb, thick, dashed] ({\iyz * \radiusats}, {\iyx * \radiusats}) arc[start angle={atan2(\iyx, \iyz)}, delta angle=180, radius=\radiusats cm];
			\draw[color=colorb, thick] ({-\iyz * \radiusats}, {-\iyx * \radiusats}) arc[start angle={atan2(-\iyx, -\iyz)}, delta angle=180, radius=\radiusats cm];
		\end{scope}
		
		% cut 3
		\begin{scope}[canvas is zx plane at y=\s]
			\filldraw[color=colorc, thick, draw opacity=1, fill opacity=0.1] (-\margin, -\margin) -- (\margin, -\margin) -- (\margin, \margin) -- (-\margin, \margin) -- cycle;
			\draw[color=colorc, thick, dashed] ({\iyz * \radiusats}, {\iyx * \radiusats}) arc[start angle={atan2(\iyx, \iyz)}, delta angle=180, radius=\radiusats cm];
			\draw[color=colorc, thick] ({-\iyz * \radiusats}, {-\iyx * \radiusats}) arc[start angle={atan2(-\iyx, -\iyz)}, delta angle=180, radius=\radiusats cm];
		\end{scope}
		
		% meridians and equator
		\begin{scope}[canvas is yz plane at x=0]
			\draw (\ixy, \ixz) arc[start angle={atan2(\ixz, \ixy)}, delta angle=180, radius=1 cm];
			\draw[dashed] (-\ixy, -\ixz) arc[start angle={atan2(-\ixz, -\ixy)}, delta angle=180, radius=1 cm];
		\end{scope}
		\begin{scope}[canvas is zx plane at y=0]
			\draw[dashed] (\iyz, \iyx) arc[start angle={atan2(\iyx, \iyz)}, delta angle=180, radius=1 cm];
			\draw (-\iyz, -\iyx) arc[start angle={atan2(-\iyx, -\iyz)}, delta angle=180, radius=1 cm];
		\end{scope}
		\begin{scope}[canvas is xy plane at z=0]
			\draw[dashed] (\izx, \izy) arc[start angle={atan2(\izy, \izx)}, delta angle=180, radius=1 cm];
			\draw (-\izx, -\izy) arc[start angle={atan2(-\izy, -\izx)}, delta angle=180, radius=1 cm];
		\end{scope}
		
		% empty case
		\draw[color=colora] (3.75*\xx+\xy, 3.75*\yx+\yy, 3.75*\zx+\zy) node {$s > b$};
		
		% northern case
		\begin{scope}[shift={(4.5*\xx-0.2*\xy, 4.5*\yx-0.2*\yy, 4.5*\zx-0.2*\zy)}, scale=0.5]
			\draw[color=colorb] (-3.5, +1.3, 0) node {$\frac{1}{2} ( a + b ) < s < b$};
			
			% north pole
			\shadedraw (0, \radius, 0) circle (0.03 cm) node[above] {$\mathsf{b}$};
			
			% Bloch sphere
			\draw (\xy, \yy, \zy) arc[start angle=90, end angle={acos(\radiusats*\iyx*\xx - \s*\yx + \radiusats*\iyz*\zx)}, radius=1 cm];
			\draw (\xy, \yy, \zy) arc[start angle=90, end angle={acos(-\radiusats*\iyx*\xx - \s*\yx - \radiusats*\iyz*\zx)}, radius=1 cm];
			
			% cut
			\begin{scope}[canvas is zx plane at y=-\s]
				\draw[color=colorb, thick, dashed] ({\iyz * \radiusats}, {\iyx * \radiusats}) arc[start angle={atan2(\iyx, \iyz)}, delta angle=180, radius=\radiusats cm];
				\draw[color=colorb, thick] ({-\iyz * \radiusats}, {-\iyx * \radiusats}) arc[start angle={atan2(-\iyx, -\iyz)}, delta angle=180, radius=\radiusats cm];
			\end{scope}
			
			% meridians
			\begin{scope}[canvas is yz plane at x=0]
				\draw (\ixy, \ixz) arc[start angle={atan2(\ixz, \ixy)}, end angle=90-\minstheta, radius=1 cm];
				\draw[dashed] (\ixy, \ixz) arc[start angle={atan2(\ixz, \ixy)}, end angle=\minstheta-90, radius=1 cm];
			\end{scope}
			\begin{scope}[canvas is xy plane at z=0]
				\draw (-\izx, -\izy) arc[start angle={atan2(-\izy, -\izx)}, end angle=180-\minstheta, radius=1 cm];
				\draw[dashed] (-\izx, -\izy) arc[start angle={atan2(-\izy, -\izx)}, end angle=\minstheta, radius=1 cm];
			\end{scope}
		\end{scope}

		% southern case
		\begin{scope}[shift={(3*\xx-\xy, 3*\yx-\yy, 3*\zx-\zy)}, scale=0.5]
			\draw[color=colorc] (3.5, -0.5, 0) node {$s < \frac{1}{2} ( a + b )$};
			
			% north pole
			\shadedraw (0, 0, 0) circle (0.03 cm) node[below] {$\mathsf{o}$};
			\shadedraw (0, \radius, 0) circle (0.03 cm) node[above] {$\mathsf{b}$};
			
			% Bloch sphere
			\draw (\xy, \yy, \zy) arc[start angle=90, end angle={-acos(\radiusats*\iyx*\xx + \s*\yx + \radiusats*\iyz*\zx)}, radius=1 cm];
			\draw (\xy, \yy, \zy) arc[start angle=90, end angle={360-acos(-\radiusats*\iyx*\xx + \s*\yx - \radiusats*\iyz*\zx)}, radius=1 cm];
			% cut
			\begin{scope}[canvas is zx plane at y=\s]
				\draw[color=colorc, thick, dashed] ({\iyz * \radiusats}, {\iyx * \radiusats}) arc[start angle={atan2(\iyx, \iyz)}, delta angle=180, radius=\radiusats cm];
				\draw[color=colorc, thick] ({-\iyz * \radiusats}, {-\iyx * \radiusats}) arc[start angle={atan2(-\iyx, -\iyz)}, delta angle=180, radius=\radiusats cm];
			\end{scope}
			
			% meridians and equator
			\begin{scope}[canvas is yz plane at x=0]
				\draw (\ixy, \ixz) arc[start angle={atan2(\ixz, \ixy)}, end angle=90-\stheta, radius=1 cm];
				\draw[dashed] (\ixy, \ixz) arc[start angle={atan2(\ixz, \ixy)}, end angle=\stheta-90, radius=1 cm];
			\end{scope}
			\begin{scope}[canvas is zx plane at y=0]
				\draw[dashed] (\iyz, \iyx) arc[start angle={atan2(\iyx, \iyz)}, delta angle=180, radius=1 cm];
				\draw (-\iyz, -\iyx) arc[start angle={atan2(-\iyx, -\iyz)}, delta angle=180, radius=1 cm];
			\end{scope}
			\begin{scope}[canvas is xy plane at z=0]
				\draw (-\izx, -\izy) arc[start angle={atan2(-\izy, -\izx)}, end angle=180-\stheta, radius=1 cm];
				\draw[dashed] (-\izx, -\izy) arc[start angle={atan2(-\izy, -\izx)}, end angle=\stheta, radius=1 cm];
			\end{scope}
		\end{scope}
	\end{tikzpicture}
	\caption{For Choquet integral calculation, sets $A(s)$ of all pure states for which the statistical variable is greater than $s$. On the left, the Bloch ball with planes at constant $s$. On the right, the top case is when the value is greater than the maximum (empty set); the middle case is when the value is between the target point $\ens[o]$ and the maximum; the third case is when the value is below the target point.}\label{fig_ChoquetCuts}
\end{figure}

The ensemble space $\Ens$ for a two-state quantum system is the Bloch ball. The pure states, the extreme points, are on the surface $X$ and the internal points correspond to mixed states. The center point $\ens[o]$ is the maximally mixed state. Since $\ens[o]$ can always be obtained by the equal mixture of two opposite points (i.e.~two orthogonal states), the fraction capacity for every singleton of an extreme point $\psi$ is $\frcap_{\ens[o]}(\{\psi\}) =\frac{1}{2}$. A statistical variable $F$ (i.e.~a quantum observable), is an affine function of the ensemble space. This means that two opposite points on the surface $\ens[a]$ and $\ens[b]$ (i.e.~the eigenstates) will take the minimum and maximum values $a = F(\ens[a])$ and $b = F(\ens[b])$. For simplicity, we can picture $\ens[a]$ and $\ens[b]$ as the extreme points of the vertical axis and we will assume $0<a<b$.

The Choquet integral is defined as 
\begin{equation}
	\begin{aligned}
(C)\int F d\frcap_{\ens[o]} := &\int_{-\infty}^0
(\frcap_{\ens[o]} (\{x | F (x) \geq s\})-\frcap_{\ens[o]}(X))\, ds \\
&+
\int^\infty_0
\frcap_{\ens[o]} (\{x | F (x) \geq s\})\, ds.
	\end{aligned}
\end{equation}
The set $A(s) = \{x | F (x) \geq s\}$ corresponds to all the points of $X$, the surface of the sphere, that have a value of $F$ greater than a given $s$. Since $F$ is an affine function, this corresponds to the set of points above a given height. So, all we need to do to calculate the integral is calculate the fraction capacity for sets of spheres $A(s)$ where the bottom is cut off at all possible values. See Fig. \ref{fig_ChoquetCuts}.

If $s\leq \frac{1}{2} F(\ens[a]) + \frac{1}{2} F(\ens[b])$, we are cutting the sphere at the midpoint or below. This means that $A(s)$ will include two opposite points. Since $\ens[o]$ can be expressed as the mixture of two opposite points, $\ens[o]$ is in the hull of $A(s)$ and therefore $\frcap_{\ens[o]}(A(s)) = 1$. Conversely, if $s > F(\ens[b])$, $A(s) = \emptyset$, which means $\frcap_{\ens[o]}(A(s)) = 0$. The non-trivial case, then, is when $\frac{1}{2} F(\ens[a]) + \frac{1}{2} F(\ens[b]) < s \leq F(\ens[b])$.

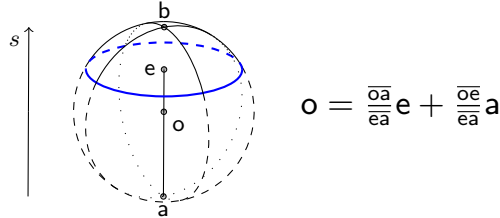
\begin{figure}[h]
	\centering
	
	% perspective
	\newcommand{\yaw}{11}
	\newcommand{\pitch}{28}
	\newcommand{\roll}{23}
	
	% transformation matrix = components of data axes in canvas coordinates
	\pgfmathsetmacro{\xx}{cos(\yaw)*cos(\pitch)}
	\pgfmathsetmacro{\xy}{sin(\yaw)*cos(\pitch)}
	\pgfmathsetmacro{\xz}{-sin(\pitch)}

	\pgfmathsetmacro{\yx}{cos(\yaw)*sin(\pitch)*sin(\roll) - sin(\yaw)*cos(\roll)}
	\pgfmathsetmacro{\yy}{sin(\yaw)*sin(\pitch)*sin(\roll) + cos(\yaw)*cos(\roll)}
	\pgfmathsetmacro{\yz}{cos(\pitch)*sin(\roll)}
	
	\pgfmathsetmacro{\zx}{cos(\yaw)*sin(\pitch)*cos(\roll) + sin(\yaw)*sin(\roll)}
	\pgfmathsetmacro{\zy}{sin(\yaw)*sin(\pitch)*cos(\roll) - cos(\yaw)*sin(\roll)}
	\pgfmathsetmacro{\zz}{cos(\pitch)*cos(\roll)}

	% intersection of yz grand circle with horizon
	\pgfmathsetmacro{\ixx}{0}
	\pgfmathsetmacro{\ixy}{\zz / sqrt(\zz*\zz + \yz*\yz)}
	\pgfmathsetmacro{\ixz}{-\yz / sqrt(\zz*\zz + \yz*\yz)}
	% intersection of zx grand circle with horizon
	\pgfmathsetmacro{\iyx}{\zz / sqrt(\zz*\zz + \xz*\xz)}
	\pgfmathsetmacro{\iyy}{0}
	\pgfmathsetmacro{\iyz}{-\xz / sqrt(\zz*\zz + \xz*\xz)}
	% intersection of xy grand circle with horizon
	\pgfmathsetmacro{\izx}{-\yz / sqrt(\xz*\xz + \yz*\yz)}
	\pgfmathsetmacro{\izy}{\xz / sqrt(\xz*\xz + \yz*\yz)}
	\pgfmathsetmacro{\izz}{0}

	% dimensions
	\newcommand{\s}{0.5}
	\pgfmathsetmacro{\radiusats}{sqrt(1 - \s*\s)}
	\pgfmathsetmacro{\stheta}{asin(\s)}
	\pgfmathsetmacro{\minstheta}{asin(-\s)}
	
	\newcommand{\radius}{1}
	\newcommand{\margin}{1.3}
	
	\begin{tikzpicture}[x={(\xx cm, \xy cm)}, y={(\yx cm, \yy cm)}, z={(\zx cm, \zy cm)}, scale=1.2]
		% Bloch sphere (outline drawn in canvas plane)
		\draw (\xy, \yy, \zy) arc[start angle=90, end angle={acos(\radiusats*\iyx*\xx - \s*\yx + \radiusats*\iyz*\zx)}, radius=1 cm] coordinate (A);
%		\draw[dash dot, red] (A) arc[start angle={acos(\radiusats*\iyx*\xx - \s*\yx + \radiusats*\iyz*\zx)}, end angle=-90, radius=1 cm];
		\draw (\xy, \yy, \zy) arc[start angle=90, end angle={acos(-\radiusats*\iyx*\xx - \s*\yx - \radiusats*\iyz*\zx)}, radius=1 cm] coordinate (B);
%		\draw[dash dot, green] (B) arc[start angle={acos(-\radiusats*\iyx*\xx - \s*\yx - \radiusats*\iyz*\zx)}, end angle=270, radius=1 cm];
		\draw[dashed] (A) arc[start angle={acos(\radiusats*\iyx*\xx - \s*\yx + \radiusats*\iyz*\zx)}, end angle={acos(-\radiusats*\iyx*\xx - \s*\yx - \radiusats*\iyz*\zx)-360}, radius=1 cm];

		% special points
		\shadedraw (0, 0, 0) circle (0.03 cm) node[below right] {$\mathsf{o}$};
		\shadedraw (0, \radius, 0) circle (0.03 cm) node[above] {$\mathsf{b}$};
		\shadedraw (0, -\radius, 0) circle (0.03cm) node[below] {$\mathsf{a}$};
		\shadedraw (0, \s, 0) circle (0.03cm) node[left] {$\mathsf{e}$};
		
		% convex combination
		\draw (0, -\radius, 0) -- (0, \s, 0);
		\draw (2.75*\xx, 2.25*\yx, 2.25*\zx) node[scale=1.4] {$\ens[o] = \frac{\overline{\ens[o]\ens[a]}}{\overline{\ens[e]\ens[a]}} \ens[e] + \frac{\overline{\ens[o]\ens[e]}}{\overline{\ens[e]\ens[a]}} \ens[a]$};
		
		% s axis
		\begin{scope}[shift={(-1.5*\xx, -1.5*\yx, -1.5*\zx)}]
			\draw (0, -1, 0) -- (0, 0, 0);
			\draw[->] (0, 0, 0) -- (0, 1, 0) node[below left] {$s$};
		\end{scope}
		
		% cut
		\begin{scope}[canvas is zx plane at y=\s]
%			\filldraw[color=blue, thick, draw opacity=1, fill opacity=0.1] (-\margin, -\margin) -- (\margin, -\margin) -- (\margin, \margin) -- (-\margin, \margin) -- cycle;
			\draw[blue, thick, dashed] ({\iyz * \radiusats}, {\iyx * \radiusats}) arc[start angle={atan2(\iyx, \iyz)}, delta angle=180, radius=\radiusats cm];
			\draw[blue, thick] ({-\iyz * \radiusats}, {-\iyx * \radiusats}) arc[start angle={atan2(-\iyx, -\iyz)}, delta angle=180, radius=\radiusats cm];
		\end{scope}
		
		% meridians
		\begin{scope}[canvas is yz plane at x=0]
			% foreground above s
			\draw (\ixy, \ixz) arc[start angle={atan2(\ixz, \ixy)}, end angle=90-\stheta, radius=1 cm] coordinate (C);
			% foreground below s
			\draw[dashed] (C) arc[start angle=90-\stheta, end angle={atan2(-\ixz, -\ixy)}, radius=1 cm] coordinate (D);
			% background below s
			\draw[loosely dotted] (D) arc[start angle={atan2(-\ixz, -\ixy)}, end angle=\stheta+270, radius=1 cm];
			% background above s
			\draw[densely dotted] (\ixy, \ixz) arc[start angle={atan2(\ixz, \ixy)}, end angle=\stheta-90, radius=1 cm];
		\end{scope}
		\begin{scope}[canvas is xy plane at z=0]
			% foreground above s
			\draw (-\izx, -\izy) arc[start angle={atan2(-\izy, -\izx)}, end angle=180-\stheta, radius=1 cm] coordinate (E);
			% foreground below s
			\draw[dashed] (E) arc[start angle=180-\stheta, end angle={atan2(\izy, \izx)+360}, radius=1 cm] coordinate (F);
			% background below s
			\draw[loosely dotted] (F) arc[start angle={atan2(\izy, \izx)}, end angle=\stheta, radius=1 cm];
			% background above s
			\draw[densely dotted] (-\izx, -\izy) arc[start angle={atan2(-\izy, -\izx)}, end angle=\stheta, radius=1 cm];
		\end{scope}
	\end{tikzpicture}
	\caption{Geometric representation of the element $\ens[e]$ in $A(s)$ with the largest fraction of $\ens[o]$ when $\frac{1}{2} F(\ens[a]) + \frac{1}{2} F(\ens[b]) < s \leq F(\ens[b])$.}\label{fig_ChoquetTop}
\end{figure}

In this case, the hull of $A(s)$ is the top spherical cap. We need to consider all possible convex combinations $\ens[o] = p \ens + \bar{p} \ens[c]$ where $\ens \in A(s)$ is in the spherical cap and $\ens[c]$ is not, and find the one with the highest $p$. Since, for combinations along the same direction, $p$ increases as $\ens$ gets closer to $\ens[o]$ and as $\ens[c]$ gets further, $\ens[c]$ will be an extreme point on the lower hemisphere while $\ens$ will be on the disc at constant $s$, the lower side of the cap. Also note that the distance to $\ens[o]$ is the same for all extreme points, therefore, to maximize $p$, we simply need to minimize the distance between $\ens[o]$ and $\ens$. Therefore $\ens$ must be the intersection between the vertical axis and the disk, and $\ens[c]$ will coincide with $\ens[a]$ as shown in Fig. \ref{fig_ChoquetTop}. The coefficient $p$ will be given by the ratio of the distances $\overline{\ens[o]\ens[a]}$ and $\overline{\ens\ens[a]}$. Since $F$ is an affine functional, we have
\begin{equation}
	\frac{\overline{\ens[o]\ens[a]}}{\overline{\ens\ens[a]}} = \frac{F(\ens[o])-F(\ens[a])}{F(\ens) - F(\ens[a])} = \frac{1}{2} \frac{b - a}{s-a}.
\end{equation}

To recap, as shown in Fig. \ref{fig_ChoquetIntegrand} we have
\begin{equation}
	\frcap_{\ens[o]}(A(s))=
	\begin{cases}
		1 & \text{if } s\leq \frac{1}{2} a + \frac{1}{2} b\\
		\frac{1}{2} \frac{b - a}{s-a} & \text{if } \frac{1}{2} a + \frac{1}{2} b < s \leq b\\
		0 & \text{if } s > b.
	\end{cases}
\end{equation}
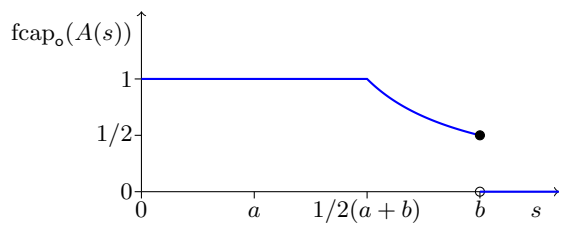
\begin{figure}[H]
	\centering
	\newcommand{\ticklength}{0.03}
	\begin{tikzpicture}[scale=3, xticklabel/.style={anchor=base, inner sep=1pt, yshift=-2.5ex}]
		% axes
		\draw[->] (0, 0) -- (1.85, 0);
		\draw (0, 0) node[left] {$0$} -- +(-\ticklength, 0) (0, 0.25) node[left] {$1/2$} -- +(-\ticklength, 0) (0, 0.5) node[left] {$1$} -- +(-\ticklength, 0);
		\draw[->] (0, 0) -- (0, 0.8);
		\draw (0, 0) node[xticklabel] {$0$} -- +(0, -\ticklength) (0.5, 0) node[xticklabel] {$a$} -- +(0, -\ticklength) (1, 0) node[xticklabel] {$1/2(a+b)$} -- +(0, -\ticklength) (1.5, 0) node[xticklabel] {$b$} -- +(0, -\ticklength);
		% cases s <= 1/2(a+b) and 1/2(a+b) < s <= b
		\draw[thick, blue, domain=1:1.5, samples=200] (0, 0.5) --plot (\x, {0.25/(\x-0.5)});
		% mark discontinuity
		\filldraw (1.5, 0.25) circle (0.02cm);
		\draw (1.5, 0) circle (0.02cm);
		% case s > b
		\draw[thick, blue] (1.5, 0) -- (1.85, 0);
        \draw (0,0.7) node[left] {$\frcap_{\ens[o]}(A(s))$};
        \draw (1.75,0) node[xticklabel] {$s$};
	\end{tikzpicture}
	\caption{The integrand $\frcap_{\ens[o]}(A(s))$ as a function of $s$.}\label{fig_ChoquetIntegrand}
\end{figure}
We can now calculate the integral
\begin{equation}
	\begin{aligned}
		(C)&\int F d\frcap_{\ens[o]} := 
		\int_{-\infty}^0
		(\frcap_{\ens[o]} (A(s))-\frcap_{\ens[o]}(X))\, ds +
		\int^\infty_0
		\frcap_{\ens[o]} (A(s))\, ds \\
		&= \int_{-\infty}^0 (1 - 1)ds + \int_{0}^{\frac{1}{2} a + \frac{1}{2} b}  1 ds + \int_{\frac{1}{2} a + \frac{1}{2} b}^{b}  \frac{1}{2} \frac{b - a}{s-a} ds + \int_{b}^{\infty} 0 ds \\
		&=\left[s\right]_{0}^{\frac{1}{2} a + \frac{1}{2} b}
		+ \frac{1}{2}\left(b - a\right)\left[\ln (s - a)\right]_{\frac{1}{2} a + \frac{1}{2} b}^{b} \\
		&= \frac{1}{2} a + \frac{1}{2} b + \frac{1}{2}\left(b - a\right)\left[\ln (b - a) - \ln \left(\frac{1}{2} a + \frac{1}{2} b - a\right) \right] \\
		&= \frac{1}{2} a + \frac{1}{2} b + \frac{1}{2}\left(b - a\right)\ln (2)
	\end{aligned}
\end{equation}

Note that, since $\ens[o]$ is the center of the ball, $F(\ens[o]) = \frac{1}{2} a + \frac{1}{2} b \neq \frac{1}{2} a + \frac{1}{2} b + \frac{1}{2}\left(b - a\right)\ln (2)$. This means that the Choquet integral of the statistical variable with respect to the fraction capacity does not recover the expectation value.

\section{Conclusion}

While representation of ensembles though a non-additive probability measure over pure states seems well-founded, the corresponding notion of non-additive integral does not correspond to the most commonly used ones. We therefore welcome guidance from experts in the field of non-additive/fuzzy measures.

\begin{credits}
\subsubsection{\ackname} This paper was made possible through the support of grant \#62847 from the John Templeton Foundation. It is part of the research program Assumptions of Physics \cite{aop-book}, which aims to identify a handful of physical principles from which the basic laws can be rigorously derived. Figures are courtesy of Assumptions of Physics. We thank Alexander Wilce, Vicen\c{c} Torra, Zuzana Ontkovi\v{c}ov\'{a} and Michel Grabisch for useful pointers and discussion.

\subsubsection{\discintname}
No competing interests.
\end{credits}
\bibliographystyle{splncs04}
\bibliography{bibliography}
\end{document}